\documentclass{aa}

\usepackage[varg]{txfonts}
\usepackage{graphicx}
\usepackage{subfigure} 
\usepackage{color} 
\usepackage{natbib}

\bibpunct{(}{)}{;}{a}{}{,}
\begin{document}

\title{Elliptical instability of compressible flow in ellipsoids}
\author{N. Clausen\inst{\ref{inst1}}\and A. Tilgner\inst{\ref{inst1}}}
\institute{Institute of Geophysics, University of G\"ottingen, Friedrich-Hund-Platz 1, 37077 G\"ottingen, Germany \\ \email{[niels.clausen;andreas.tilgner]@geo.physik.uni-goettingen.de}\label{inst1}}
\date{Received xxx,Accepted xxx}

\abstract
{Elliptical instability is due to a parametric resonance of two inertial modes in a fluid velocity field with elliptical streamlines. This flow is a simple model of the motion in a tidally deformed, rotating body. Elliptical instability typically leads to three-dimensional turbulence. The associated turbulent dissipation together with the dissipation of the large scale mode may be important for the synchronization process in stellar and planetary binary systems.}
{In order to determine the influence of the compressibility on the stability limits of tidal flows in stars or planets, we calculate the growth rates of perturbations in flows with elliptical streamlines within ellipsoidal boundaries of small ellipticity. In addition, the influence of the orbiting frequency of the tidal perturber $\Omega_P$ and the viscosity of the fluid are taken into account.
%
 }{We studied the linear stability of the flow to determine the growth rates. We solved the Euler equation and the continuity equation. The viscosity was introduced heuristically in our calculations. We assumed a power law for the radial dependence of the background density. Together with the use of the anelastic approximation, this enabled us to use semi-analytical methods to solve the equations. 
}{ It is found that the growth rate of a certain mode combination depends on the compressibility. However, the influence of the compressibility is negligible for the growth rate maximized over all possible modes if viscous bulk damping effects can be neglected. The growth rate maximized over all possible modes determines the stability of the flow. The stability limit for the compressible fluid confined to an ellipsoid is the same as for incompressible fluid in an unbounded domain. Depending on the ratio $\Omega_P/\Omega_{F}$, with $\Omega_{F}$ the spin rate of the central object in the frame of the rotating tidal perturber, certain pairs of modes resonate with each other. The size of the bulk damping term depends on the modes which resonate with each other. Therefore the growth rate of the viscous flow depends on the compressibility. 
Estimates for the stability limit in viscous fluids are given.
}{}
\maketitle

\section{Introduction}

The study of tidal friction has a long history in geo- and astrophysics (e.g.,
 \citealt{darwin1879}). The basic picture of a tidally deformed fluid body is
that of an ellipsoid with the long axis pointing towards the tidal companion,
and in which the fluid rotates with respect to the major axes of the ellipsoid, so
that each fluid particle follows an elliptical streamline. It was
discovered in the mid 1970s that flows of constant vorticity with elliptical
streamlines are prone to a hydrodynamic instability, the so-called elliptical
instability (for a review, see \citealt{kerswell2002}). 
One expects to find vastly different dissipation rates in laminar
and in unstable turbulent flows. A determination of the stability limit of the
basic tidal flow is thus an indispensable prerequisite to any computation of
tidal friction. In this paper, we compute the stability limit within a simple
model of tidal flow taking into account the compressibility of a stellar or
planetary envelope.

Fluids in solid body rotation (i.e., flows with circular streamlines) support a
type of waves known as inertial waves in which the Coriolis force acts as a
restoring force. In elliptical instability, the finite ellipticity couples
inertial modes in pairs and can lead to instability if certain resonance
conditions are met. The same phenomenon is often called a triad resonance, the
triad being formed by the two inertial waves and the basic, elliptically
deformed flow which is itself counted as third wave.

In this paper, we calculate the growth rate of the elliptical instability in a
slightly deformed sphere. Incompressible flows have already been investigated in
this geometry (\citet{lacaze2003}, \citet{lebars2010}). Elliptical
instability can be studied in its simplest form in an unbounded domain
(\citealt{bayly1986,landman1987,waleffe1990,miyazaki1992,miyazaki1993,miyazaki1995}).
Another convenient geometry is an elliptically deformed cylinder
\citep{gledzer1975,malkus1989,eloy2003}, and other studies exist for deformed
spheroids \citep{gledzer1977,kerswell1994,cebron2010} and spherical shells
\citep{aldridge1997,seyed2000,cebron2012}.

Several additional ingredients have been added to the problem, such as
stratification (\citealt{miyazaki1992,miyazaki1993,guimbard2010}), magnetic fields
\citep{kerswell1994,lacaze2006,herreman2009,cebron2012b}, rotation of the elliptical perturbation
\citep{craik1989,gledzer1992,miyazaki1993,miyazaki1995,seyed2000,lebars2010},
and viscosity \citep{landman1987,kerswell1994,lacaze2003,lebars2010}. 
The importance of the elliptical instability for tidal dissipation is examined in a recent paper by \citet{barker2013}. They studied the non-linear evolution of the elliptical instability. They performed three-dimensional hydrodynamical simulations of a box with periodic boundary conditions with a base flow such that this box can be considered as a small patch of a tidally deformed fluid in a planet or a star. They found that for the astrophysically relevant values of the ellipticity the wave driving mechanism is not sustained permanently because of the presence of strong columnar vortices whose presence effectively suppresses the driving mechanism. This leads to less dissipation compared to the case with a sustained wave driving mechanism, but it is certainly higher compared to case with a stable flow. A strong stratification can prevent the process of re-laminarization as found in an experiment with a stratified fluid in a rotating cylinder by \citet{guimbard2010}. The same holds for the instability in the  presence of a weak magnetic field (see \citealt{barker2013b}, a companion paper to \citealt{barker2013}). \citet{barker2013b} found that this field prevents the vortices from forming. They calculated the dissipation and conclude that the inferred tidal dissipation is potentially important at short orbital periods. In this companion paper they also neglect the effects of a realistic geometry and the additional presence of turbulent convection, both of which possibly enhance the dissipation such that the instability becomes important more generally.

The influence of a temperature-gradient on the elliptical instability in a triaxial ellipsoid was examined by \citet{cebron2010b} with numerical methods. 
They found that the growth rate of the elliptical instability is significantly enhanced by a thermal stratification and that in a convective flow the elliptical instability can still grow, but with a reduced growth rate. They were not able to reach the regime of very large Reynolds numbers, where the results of \citet{fabijonas2003} predict an increased tidal destabilization.
For stars the case of a bi-layer flow with a stable stratified and a convective region is interesting.
In the paper by \citet{cebron2010b} it was shown that even in such a flow the instability can grow over the whole fluid. 

In this paper, we will include the rotation of the perturbation, viscous dissipation,
and most importantly, compressibility. The last has been neglected in nearly
all other publications on the subject with the exception of the work by
\citet{cebron2013} who numerically simulated one particular set of parameters.
In contrast, we use a semi-analytical method to obtain a broad overview. The most
important issue this study will address is how the compressibility of the flow
influences growth rates of the elliptical instability. Surprisingly, we will
find that the structure of the most unstable modes in ideal fluids depend on the
compressibility, but the maximum growth rate is independent of
compressibility. Section 2 collects all the formulas describing the model and Sect. 3
explains their numerical implementation. The results are presented in Sect. 4.

\section{Mathematical formulation of the model}

We consider the equatorial tide raised on a central body by a tidal perturber or
perturbing body. We choose a frame of reference in which the perturbing body is at
rest and use a Cartesian coordinate system $x,y,z$ with its origin at the center
of the central body, its $z-$axis directed
along the rotation axis of the reference system and the $x-$axis is pointing
towards the perturbing body. This reference frame rotates at
rate $\mathbf{\Omega}_P = \Omega_P \mathbf{\hat{z}}$ (hats denote unit vectors)
relative to inertial space. In this frame, the central body rotates about the
$z-$axis with angular velocity $\frac{1}{2} (\frac{b}{a}+\frac{a}{b}) \Omega_F$
within an ellipsoid with semi-major axes $a,b,c$ and the surface
\begin{equation}
 \left(\frac x a\right)^2+\left(\frac y b\right)^2+\left(\frac z c\right)^2=1,
\label{surf_S}
\end{equation}
so that the motion of the central body within the chosen frame of reference is
given by the velocity field $\mathbf{u_0}$:
\begin{equation}
 \mathbf{u_0}=\Omega_F\left(-\frac{a}{b}y\mathbf{\hat{x}}+\frac b a x
\mathbf{\hat{y}}\right).
\end{equation}
In order to determine the stability of this flow, we start from the full
Euler equation
\begin{equation}
\partial_t \mathbf v + (\mathbf v \cdot \nabla) \mathbf v + 2 \mathbf{\Omega}_P
\times \mathbf v = -
\frac{1}{\varrho} \nabla P + \nabla \Phi_{self} + \nabla \Phi_P
\end{equation}
\begin{equation}
\partial_t \varrho + \nabla \cdot (\varrho \mathbf v) = 0,
\end{equation}
where $\mathbf v$ stands for the velocity, $\varrho$ the density, $P$ the pressure,
and $\Phi_{self}$ for the potential terms created by the central body itself
(gravitational and centrifugal), whereas $\Phi_P$ is the perturbing tidal
potential. We will assume $\mathbf u_0$ to be a stationary solution of the above
equations for suitable density profiles and potentials $\rho_0$ and $\Phi_0$:
\begin{equation}
(\mathbf u_0 \cdot \nabla) \mathbf u_0 + 2 \mathbf{\Omega}_P
\times \mathbf u_0 = -
\frac{1}{\rho_0} \nabla p_0 + \nabla \Phi_0 + \nabla \Phi_P
\label{stationary}
\end{equation}
\begin{equation}
\nabla \cdot (\rho_0 \mathbf u_0) = 0.
\end{equation}
In order to end with a tractable problem, we will have to choose a density
profile such that the eigenmode calculation below leads to a separable equation.
This is achieved by setting
\begin{equation}
 \rho_0=\tilde \rho_0 \left(1 -
 \left(\frac x a\right)^2-\left(\frac y b\right)^2-\left(\frac z
c\right)^2\right)^\beta
\label{basicden}
\end{equation}
for arbitrary prefactors $\tilde \rho_0$ and exponents $\beta$ (\citet{wu2004}).

The question arises which beta one should use for the calculations. We must first
determine which polytropic index $n$ in the polytropic relation
\begin{equation}\label{polrel}
 P(r)=K\rho^{1+1/n}(r)
\end{equation}
is appropriate for the central body. We choose $n=3$, $n=3/2$, and $n=0$ for our calculations; 
$n=3$ is suitable for stars which are well modeled by a relativistic completely
degenerate electron gas, like relativistic white dwarfs. The same polytropic index also describes main sequence stars  
with $M\gtrsim M_{\odot}$, \citep{kippenhahn1990}.
 $n=3/2$ is appropriate for objects which are well modeled by a non relativistic
completely degenerate electron gas, like non-relativistic white dwarfs and brown
dwarfs. And $n=3/2$ is also relevant for main sequence stars with a  mass below $M\sim 0.4 M_\odot$, these are fully convective stars \citep{kippenhahn1990,chabrier2009}.
For planets, the range of $n$ is $0 \le
n\lesssim 1.5$, depending on how massive they are \citep{horedt2004}. For
Jupiter mass objects, $n=1$ is a good value \citep{chabrier2009}.
 $\rho(r)$ in the polytropic case can be obtained by solving the Lane-Emden equation. This equation can be derived through the usage of (\ref{polrel}), the equation for hydrostatic equilibrium and the Poisson equation \citep{kippenhahn1990}. The values for $\rho(r)$ in the polytropic case are taken from \citet{horedt1986}.
We determine the appropriate $\beta$ by simply fitting the power law (\ref{basicden}) to the $\rho(r)$ for the polytropic profiles.


It is easily verified that $\nabla \cdot (\rho_0 \mathbf u_0) = 0$ and
$\nabla \times \{ (\mathbf u_0 \cdot \nabla) \mathbf u_0 + 2
\mathbf{\Omega}_P\times \mathbf u_0 \} =0$. The curl of the gradient terms 
in Eq. (\ref{stationary}) is
trivially zero, and $\nabla \times \{\frac{1}{\rho_0} \nabla p_0 \}
=\frac{1}{\rho_0^2} \nabla \rho_0 \times \nabla p_0$ which is zero in a
polytropic atmosphere in hydrostatic equilibrium. $\mathbf u_0$ and $\rho_0$ as
given above are therefore solutions of the Euler equation for some
perturbing potential, albeit not necessarily for the perturbing potential of a
point mass at a finite distance. One may view $\rho_0$ either as an approximation
to the density profile of the state excited by a tidal perturber idealized as a point mass,
or as the exact profile for a perturbing potential which approximates a real
tidal potential.

We will now consider the linear stability of the ground state.
All quantities are decomposed into their value in the basic state indicated by
an index zero and a perturbation which is considered to be small:
$\varrho = \rho_0 + \rho$, $\mathbf v = \mathbf u_0 + \mathbf u$, $P=p_0+p$, and 
$\Phi_{self}=\Phi_0+\Phi$. The linearized equations are:
\begin{equation}
\partial_t \mathbf u + (\mathbf u_0 \cdot \nabla) \mathbf u +
(\mathbf u \cdot \nabla) \mathbf u_0 + 2 \mathbf{\Omega}_P \times \mathbf u
=
-\frac{1}{\rho_0} \nabla p + \frac{\rho}{\rho_0^2}\nabla p_0 + \nabla \Phi
\end{equation}
\begin{equation}
\partial_t \rho + \nabla \cdot (\rho \mathbf u_0 + \rho_0 \mathbf u) = 0.
\end{equation}
We will restrict ourselves to well mixed ground states $\rho_0,p_0$ of constant
entropy in atmospheres characterized by an adiabatic exponent $\gamma$, and
only allow perturbations of the ground state which obey the adiabatic equation of
state, so that $\nabla \rho_0 = \frac{\rho_0}{\gamma p_0} \nabla p_0$
and $p/p_0 = \gamma \rho/\rho_0$ for small deviations $\rho,p$ from the ground
state. The right hand side of the linearized Euler equation then simplifies according
to $\nabla (\frac{p}{\rho_0}) = - \frac{\rho}{\rho_0^2} \nabla p_0 +
\frac{1}{\rho_0} \nabla p$. 

The equation of continuity is next simplified by invoking the anelastic
approximation. This approximation is valid if the density deviations from
$\rho_0$ are small and if velocities are small compared with the speed of sound.
The latter condition is violated at the surface of a star or a planet where the
speed of sound tends to zero. The validity of the anelastic approximation has
already been discussed in detail in the more restricted scope of eigenmode
calculations as they will be done below. It appears that the region where the
anelastic approximation fails is too small to modify global results such as
inertial mode frequencies \citep{ivanov2010}, so that we shall adopt the anelastic
approximation from here on.

The linear stability problem is now reduced to
\begin{equation}
\partial_t \mathbf u + (\mathbf u_0 \cdot \nabla) \mathbf u +
(\mathbf u \cdot \nabla) \mathbf u_0 + 2 \mathbf{\Omega}_P \times \mathbf u
= \nabla \psi
\label{EEpsi}
\end{equation}
\begin{equation}
\nabla \cdot (\rho_0 \mathbf u) = 0
\end{equation}
with $\psi = \frac{p}{\rho_0} + \Phi$. We will solve these equations subject to
the boundary condition $\mathbf{\hat{n}} \cdot \mathbf{u}=0$ on the ellipsoidal
surface (\ref{surf_S}). These boundary conditions describe a solid wall in a
laboratory experiment more directly than the surface of an atmosphere, but the
results below suggest that the detailed choice of boundary conditions does not
matter for the main findings of this paper.

We now follow a procedure similar to that used by \citet{gledzer1992}. We first
remove dimensions by rescaling the Cartesian coordinates with their respective
semi-major axes (which maps the ellipsoidal surface on a sphere of radius 1) and
we rescale time with $\Omega_F$,
\begin{equation}
\begin{aligned}
 x^\prime&=\frac{x}{a},\quad y^\prime=\frac{y}{b},\quad
z^\prime=\frac{z}{c},\quad u^\prime=\frac{u}{\Omega_F a},\quad
v^\prime=\frac{v}{\Omega_F b},\\ 
w^\prime&=\frac{w}{\Omega_F c},\quad
t^\prime=\Omega_F t,\quad   R=\sqrt{\frac{a^2+b^2}{2}},\quad
\psi^\prime=\frac{\psi}{\Omega_F^2 R^2},\\ 
\rho^\prime&=\frac{\rho}{\tilde{\rho}_0},
\label{eq13}
\end{aligned}
\end{equation}
where $u$, $v$, and $w$ are the $x$, $y$, and $z$ components of $\mathbf u$,
respectively. 
For simplification we consider only the case $c=R=\sqrt{(a^2+b^2)/2}$. It
will become plausible below 
that a change in $c$ will
not cause substantial changes. This
restriction to certain ellipsoidal shapes will simplify the calculations below. In these
new variables, Eq. \ref{EEpsi} becomes after omitting primes
\begin{equation}
\begin{aligned}
\label{EEsphere}
\frac{\partial u}{\partial t}+x\frac{\partial u}{\partial y}-y\frac{\partial u}{\partial x}-v-2 v \Omega\frac{b}{a}&=-\frac{1}{1+\epsilon}\frac{\partial \psi}{\partial x},\\
\frac{\partial v}{\partial t}+x\frac{\partial v}{\partial y}-y\frac{\partial v}{\partial x}+u+2 u \Omega\frac{a}{b}&=-\frac{1}{1-\epsilon}\frac{\partial \psi}{\partial y},\\
\frac{\partial w}{\partial t}+x\frac{\partial w}{\partial y}-y\frac{\partial w}{\partial x}&=-\frac{\partial \psi}{\partial z}
\end{aligned}
\end{equation}
\begin{equation}
 \nabla \cdot(\rho_0 \mathbf{u})=0,
\label{conti2}
\end{equation}
with $\epsilon=(a^2-b^2)/(a^2+b^2)$ being the ellipticity 
of the boundaries in the $x,y$-plane and $\Omega=\Omega_P/\Omega_F$.

We now switch from Cartesian coordinates $(x,y,z)$ to cylindrical coordinates
$(s,\phi,z)$. In cylindrical coordinates, the above system of equations can be
expressed as
\begin{equation}
 \mathbf{M}\left(\frac{\partial \mathbf{u}}{\partial t}+ \mathbf{H} \mathbf{u} \right)+2\Omega(1-\epsilon^2)^{1/2}\mathbf\Lambda\mathbf{u}=-\nabla \psi,\quad \nabla \cdot(\rho_0\mathbf{u})=0 
\label{system}
\end{equation}
with
\begin{equation*}
\begin{aligned}
\mathbf{M}&=\mathbf{I}+\epsilon \mathbf{T}, \quad \mathbf{H}=\mathbf{I}
\frac{\partial}{\partial \phi}+2\mathbf{\Lambda},\\
\mathbf{T}&=
\begin{pmatrix}
\cos(2\phi) & -\sin(2\phi) & 0 \\
-\sin(2\phi) & -\cos(2\phi) & 0 \\
0 & 0 & 0
\end{pmatrix}
, \quad
\mathbf{\Lambda}=
\begin{pmatrix}
0 & -1 & 0 \\
1 & 0 & 0 \\
0 & 0 & 0
\end{pmatrix},
\end{aligned}
\end{equation*}
 where $\mathbf{I}$ is the identity matrix. It matters at this stage that we
restricted $c$ to $c=R$ because otherwise, the equation 
$\mathbf{M}=\mathbf{I}+\epsilon \mathbf{T}$ would contain additional terms.

We now use a Galerkin method. We seek solutions of \ref{system}
in the form
\begin{equation}
\begin{aligned}
 \mathbf{u}&=\sum_jC_j\mathbf{u_j}e^{i f t}\\
 \psi&=\sum_jC_j\psi_je^{i f t},
\end{aligned}
\label{ansatz}
\end{equation}
where the $\mathbf{u_j}$ are solutions of the unperturbed eigenvalue problem
($\epsilon=0$)
\begin{equation}
\begin{split}
 \mathbf{H}\mathbf{u_j}+\nabla\psi_j+2 \mathbf\Lambda \mathbf{u_j}\Omega=-i\omega_j\mathbf{u_j}, \quad \nabla
\cdot(\rho_0 \mathbf{u_j})=0,\\ \mathbf{\hat{n}} \cdot \mathbf{u_j}=0 \quad
\textrm{on the unit sphere.}
\end{split}
\end{equation}
This is the equation for inertial modes in a sphere. $j$ is a proxy for the
indices $(n_j,m_j,k_j)$ which characterize an inertial mode in a sphere
\citep{greenspan1968}. The variables $n$, $m$, and $k$ will be indexed by $j$
if it is necessary to distinguish different modes and will appear without an index
otherwise. $k$ is the azimuthal wavenumber, and $n$ is the spatial
degree or latitudinal wavenumber (on the surface of the sphere, the pressure
distribution is given by the spherical harmonic $Y_n^k$). The index $m$
numbers the eigenvalues for any fixed $n$ and $k$ and varies over a finite
range. These equations can be solved analytically in the incompressible case
\citep{zhang2001}. They are still separable within the anelastic approximation provided
that the density is of the form 
\begin{equation}
 \rho_0=(1-s^2-z^2)^\beta.
\end{equation}
Repeating the steps of \citet{wu2004}, one arrives at an eigenvalue problem for $\psi$
which is written as a product in the form $\psi(x_1,x_2,\phi,t)=\psi_1(x_1)\psi_2(x_2
)\zeta(\phi,t)$. $(x_1,x_2,\phi)$ are the ellipsoidal coordinates used in \citet{wu2004}
and introduced by \citet{bryan1889}. One must be careful to distinguish the index which
characterizes the different inertial modes from the index used in
Eq. (\ref{gleichung}) and (\ref{randbedingungen}) which stands for the two
coordinates $x_1$ and $x_2$. In this paragraph we omit the index which characterizes the different inertial modes.
The range of the three coordinates is $x_1 \in [\mu,1]$, $x_2 \in
[-\mu,\mu]$, with $\mu=-(\omega+k)/(2(1+\Omega))$ 
and $\phi$ is the azimuthal angle so that $\phi \in [0,2\pi]$.  $\mu$ is half the
frequency of the inertial mode with respect to the frame rotating with
$\Omega_P+\Omega_F$ and $\omega$ is the frequency with respect to the inertial
frame. 
All variables $X$ (pressure and velocity components) depend on time $t$ and on
the azimuthal angle $\phi$  through the relation $X\propto\zeta(\phi,t)=
\exp[i(k\phi-2\mu(\Omega_P+\Omega_F)t]$. We adopt the convention that $\mu\ge 0$
with $k>0$ representing a prograde mode and $k<0$ a retrograde mode. Modes with
denotation $(k,\mu)$ and $(-k,-\mu)$ are physically the same modes so that one
can restrict either $\mu$ or $k$ to positive numbers and avoid redundancy.
It is numerically more accurate to solve for the
$g_i$ defined by $g_i=\psi_i/(1-x_i^2)^{|k|/2}$, $i=1,2$. The eigenvalue problem
for the eigenvalue $l$ and the eigenfrequency $\mu$ reads 
\begin{equation}\label{gleichung}
\begin{split}
 (1-x_i^2)\frac{d^2g_i}{x_i^2}-2x_i(|k|+1)\frac{dg_i}{dx_i}+ \frac{2\beta x_i(1-x_i^2)}{x_i^2-\mu^2}\frac{dg_i}{dx_i}\\+\left[ \lambda^2-\frac{2\beta|k|x_i^2}{x_i^2-\mu^2}+\frac{2\beta k\mu}{x_i^2-\mu^2}\right]g_i=0, 
\end{split}
\end{equation}
with the boundary conditions
\begin{equation}\label{randbedingungen}
 \begin{aligned}
   \frac{dg_1}{dx_1}\bigg\vert_{x_1=1}&=\frac{\lambda^2+2\beta[(k\mu-|k|)/(1-\mu^2)]}{2(|k|+1)}g_1\vert_{x_1=1}\\
   \frac{dg_1}{dx_1}\bigg\vert_{x_1=\mu}&=-\frac{(k-|k|\mu)}{1-\mu^2}g_1\vert_{x_1=\mu}\\
   \frac{dg_2}{dx_2}\bigg\vert_{|x_2|=\mu}&=-\textrm{sgn}(x_2)\frac{(k-|k|\mu)}{1-\mu^2}g_2|_{|x_2|=\mu}
 \end{aligned}
\end{equation}
and $\lambda^2=l(l+1)-|k|(|k|+1)$. In general, both $\mu$ and $l$ have to be
determined numerically, but for $\beta=0$, $l$ is simply given by $l=n$.

Once the eigenfunctions are found, we define the scalar product
\begin{equation}
(\mathbf{u_j},\mathbf{u_i})=\int_V\mathbf{u_j}\mathbf{u_i}^*\rho_0 rdr d\phi dz
\end{equation}
which differs from the usual scalar product by the factor $\rho_0$. The
eigenvalues calculated in the anelastic approximation are orthogonal
with this scalar product. This can be proven along the same lines as the proof
for the orthogonality of inertial modes in the
incompressible case \citep{greenspan1968}.

We substitute (\ref{ansatz}) in (\ref{system}) and multiply this equation by
$\rho_0 \mathbf{u_j}$.
We obtain the system
\begin{equation}
\begin{aligned}
 (f-\omega_j)C_jN_j^2+\epsilon(f+k_j)\sum_iV_{j,i}C_i=0,\\
N_j^2=(\mathbf{u_j},\mathbf{u_j}),\quad
 V_{i,j}=(\mathbf{T}\mathbf{u_j},\mathbf{u_i}).
\label{glpert1}
\end{aligned}
\end{equation}
We assume the $\mathbf{u_j}$ to be normalized so that $N_j^2=1$. We proceed on
the assumption that $V_{j,i}\neq 0$ only for $n_j=n_i$ and $k_j=k_i\pm 2$. The
latter condition can be proven in general by noting that
the dependence of the variables on the azimuthal angle is given by
$\exp(ik\phi)$.
The first condition is less obvious. It was proven analytically by
\citet{kerswell1993} for incompressible fluids. This proof can not
simply be adapted to the compressible case because no
analytical expressions exist for the modes in this case. However, we calculated
the $V_{j,i}$ numerically for all mode combinations with indices up to $n=20$ and
$|k|=10$. In all cases, the results appeared to converge to zero as the spatial
resolution used in the integration was improved. 

Based on the above, we can infer that the elliptical instability  comes from an
interaction of the modes with azimuthal wavenumber $k$ and $k+2$, whereas the spatial
degree $n$ must be the same for the interacting modes. Modes in resonance will be written
as $(k,k+2)$ and the spatial degree $n$ will be stated separately.

Application of perturbation theory to (\ref{glpert1}) with the small parameter
$\epsilon$ leads to the inviscid growth rate $\sigma_{inv}$
\begin{equation}
\sigma_{inv}^2=-\frac{(\epsilon^2V_{i,j}^2(k_i-k_j)+\delta\omega)^2+4\epsilon^2V_{i,j}^2q_iq_j}{4(1-\epsilon^2V^2)^2}
\label{gledz1}
\end{equation}
with $\delta\omega=\omega_i-\omega_j$, $V^2_{i,j}=V_{i,j}\cdot V_{j,i}$, and
$q_j=-2\mu_j$. Eq. (\ref{gledz1}) is the same as Eq. (3.5) of
\citet{gledzer1992}. It follows from this equation that instability is possible
only if $|\delta\omega|\leq O(\epsilon)$.

Viscosity is included heuristically in our calculations by adding a damping rate
to $\sigma_{inv}$. Two different expressions are used, one for the growth rate
with free slip boundaries, $\sigma_{fs}$, and one for no slip boundaries,
$\sigma_{ns}$. The two approximations are
\begin{equation}
\sigma_{fs}=\sigma_{inv}-g_1\, n(n+1) Ek
\label{fsbc}
\end{equation}
and
\begin{equation}
\sigma_{ns}=\sigma_{inv}-g_2 |1+\Omega| \sqrt{Ek}-g_1 \ n(n+1) Ek,
\label{nsbc}
\end{equation} 
where $g_1$ and $g_2$ are constants on the order of 1 which in this paper are always chosen as
$g_1=g_2=1$, $n$ is the spatial degree of the two interacting modes and $Ek$ is the
Ekman number. Eq. (\ref{nsbc}) contains two dissipative terms. The second one corresponds
to dissipation in the bulk, whereas the first one is due to friction inside Ekman
layers at the boundaries. The first term is usually computed in the frame of
reference in which the boundaries are at rest and expressed in multiples of the
rotation rate of that frame, yielding decay rates of inertial modes of the form $g_2
\sqrt{Ek}$ with $g_2$ typically between $0.1$ and 1 \citep{greenspan1968}.
Transformed to the frame of reference and the unit of time used in Eq.
(\ref{EEsphere}), the decay rate is $g_2 |1+\Omega| \sqrt{Ek}$ with
$Ek=\nu/(|\Omega_F+\Omega_P| R^2)$ and $\nu$ the viscosity. In many standard applications
of rotating fluid mechanics, $n$ and $Ek$ are both small enough for the term in
$\sqrt{Ek}$ to dominate. For free-slip boundary conditions on the other hand, only
the bulk dissipation must be considered.
We use $n(n+1)$ in the bulk damping term, because
the velocity field of an inertial mode contains on the surface only the
spherical harmonic $Y_n^k$. Assuming this is a reasonable approximation to the
inertial mode at any radius $r$, and using 
\begin{equation}
\nabla^2 Y_n^k(\Theta,\phi)=-\frac{(n+1)n}{r^2}  Y_n^k,
\end{equation}   
one justifies the formula for $\sigma_{fs}$ \citep{lorenzani2001}.

\section{Numerical implementation}
We used numerical methods to calculate the growth rates according to
Eq. (\ref{gledz1}). We first calculate the frequencies for $\beta=0$ of inertial modes in a
sphere by solving Eq. (29) from \citet{wu2004}
\begin{equation}\label{freq}
    \frac{d P_n^k(x_1)}{dx_1}\bigg\vert_{x_1=\mu}=-\frac{k}{1-\mu^2}P_n^k(x_1) \vert_{x_1=\mu}
\end{equation}
by bisection. We use these eigenfrequencies, together with the corresponding
eigenvalues $l$ as starting values for a shooting method to solve Eq. (\ref{gleichung}).
As noted in \citet{wu2004}, ``the number of eigenmodes remains conserved when
$\beta$ varies, with a close one-to-one correspondence between modes in
different density profiles''. So we increment $\beta$ in steps of 0.1 and use
the eigenvalues of the foregoing $\beta$ as an initial point for the
shooting method. The integration of the ODE was done by use of a Runge Kutta Cash-Karp
 method of fifth order with an adaptive stepsize control.
Because the endpoints of the ODE for one variable are both singular points, we
used the method of shooting to a fitting point \citep{press1992}.

The next step is to calculate the integrals for Eq. (\ref{gledz1}). We only have
to consider the modes which satisfy the resonance condition. Through a simple
calculation the condition for the frequency detuning $\delta \omega$
can be made more precise:
Instability is possible only if  $|\delta \omega| < 4\epsilon + 2 \epsilon^2$.

Finally we want to solve the integrals $V_{i,j}$ by Gaussian quadrature.
It is not feasible to calculate these integrals  in the ellipsoidal coordinate
space $(x_1,x_2,\phi)$ because in these coordinates the limits of integration
depend on the eigenfrequencies, and for the integration of  $V_{i,j}$ we consider
two wave functions with in general two different eigenfrequencies. Therefore we perform the integration in $(s,z,\phi)$ space. The scalars $\psi_1$ and $\psi_2$ were obtained at several points $x_1$ and $x_2$ by the numerical integration of the ODE (\ref{gleichung}). The final integrations were performed using Gaussian quadrature with Gauss-Legendre abscissas and weights with a resolution of 50 points in both $s$
and $z$ directions. The integration in $\phi$ direction can be performed analytically. We know  $x_1(s,z)$ and $x_2(s,z)$ and obtain $\psi_1(x_1(s,z),x_2(s,z)) \psi_2(x_1(s,z),x_2(s,z))$. The scalars $\psi_1$ and $\psi_2$ were obtained at some points $x_1$ and $x_2$ which in general do not coincide  with the points we need for the Gaussian quadrature. Therfore we need the  scalars $\psi_1, \psi_2$ at arbitrary points $x_1,x_2$. This was accomplished by a cubic-spline interpolation.

\section{Results}

\begin{figure}
\centering
\subfigure[\label{e004sig}]{
\includegraphics[width=0.4\textwidth,]{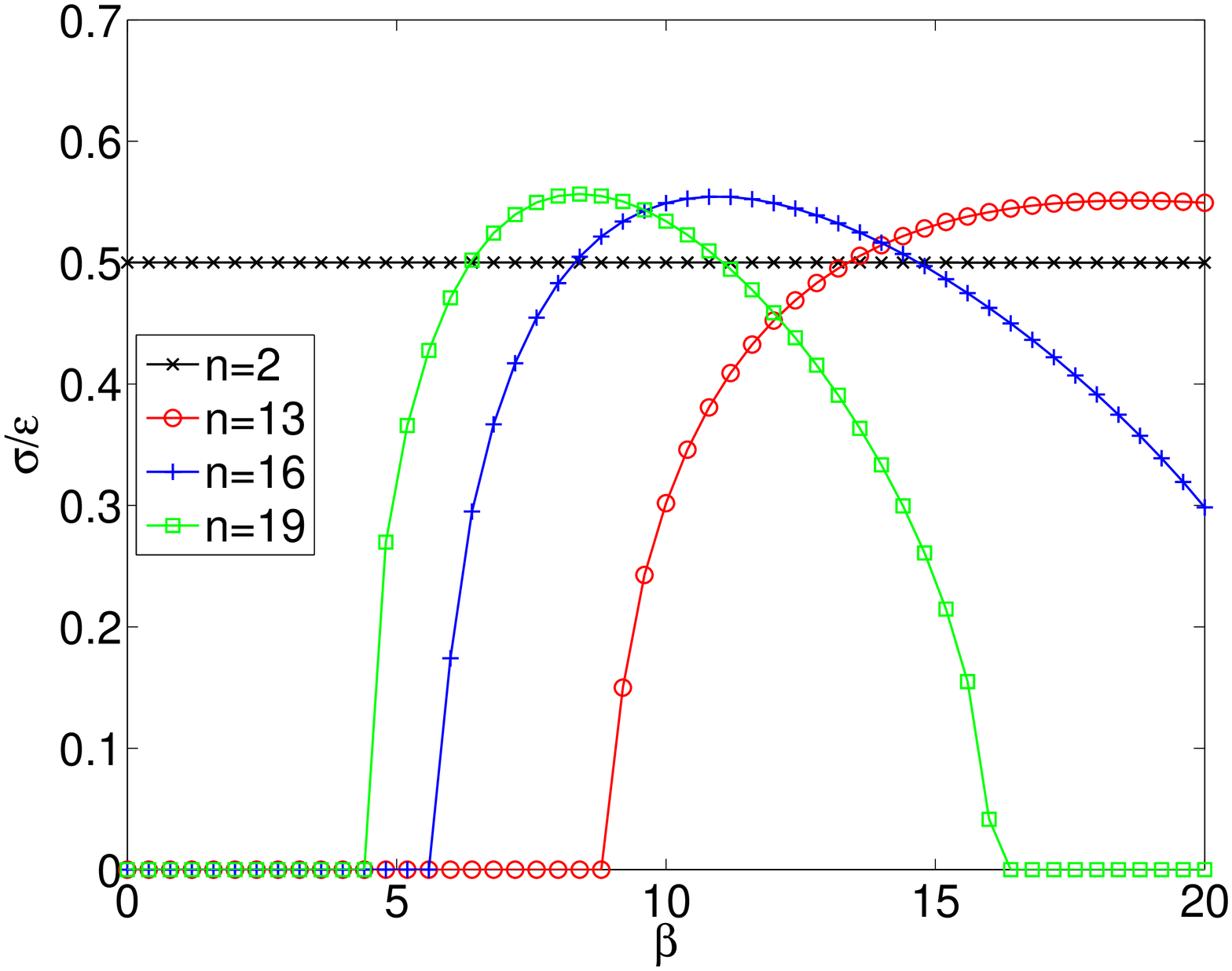}}
\subfigure[\label{dom0,04} 
]{\includegraphics[width=0.4\textwidth]{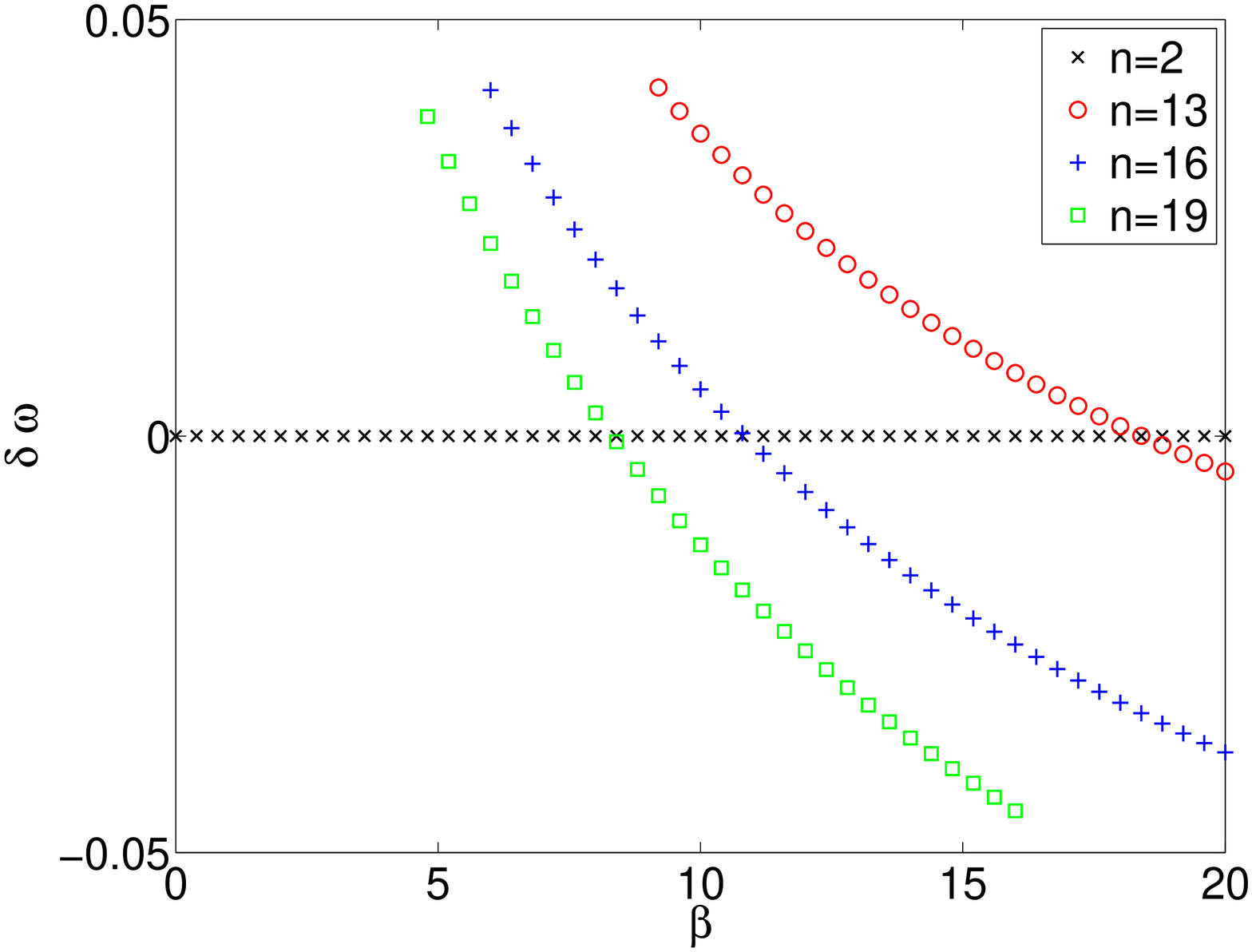}}\\
\caption{(a) Growth rates of modes (-1,1) for various $n$ as a function
of the compressibility parameter $\beta$ for $\epsilon=0.04$, $\Omega_p=0$ and $Ek=0$. (b) The corresponding $\delta \omega$.}
\label{sigbeta1}
\end{figure}

\begin{figure}
\centering
\subfigure[ \label{e0,16sig}  ]{\includegraphics[width=0.4\textwidth,]{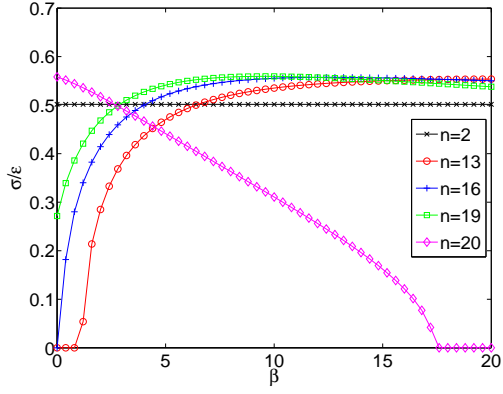}}
\subfigure[\label{dom0,16} \newline
]{\includegraphics[width=0.4\textwidth,]{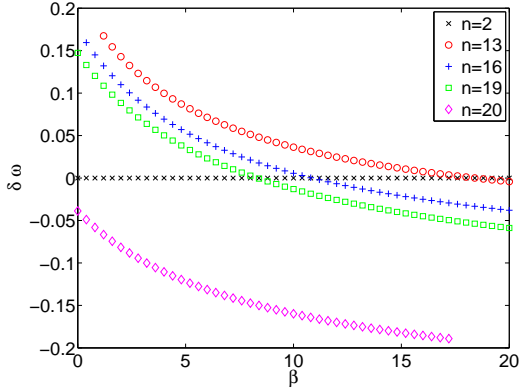}}\\
\caption{Same as fig. \ref{sigbeta1}  but for $\epsilon=0.16$.}
\label{sigbeta2}
\end{figure}

\begin{figure}
\centering
\subfigure[Modes with $kq>0$ \label{modenompos} \newline ]{\includegraphics[width=0.4\textwidth,]{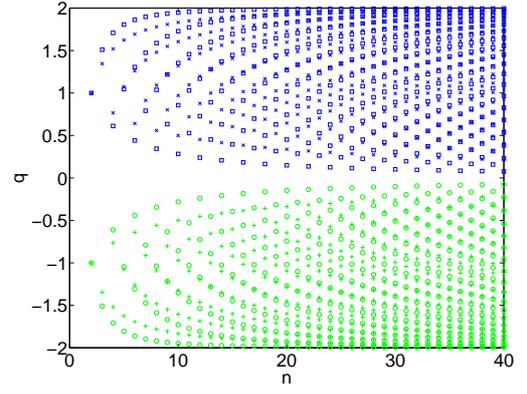}}
\subfigure[Modes with $kq<0$ \label{modenomneg}\newline ]{\includegraphics[width=0.4\textwidth]{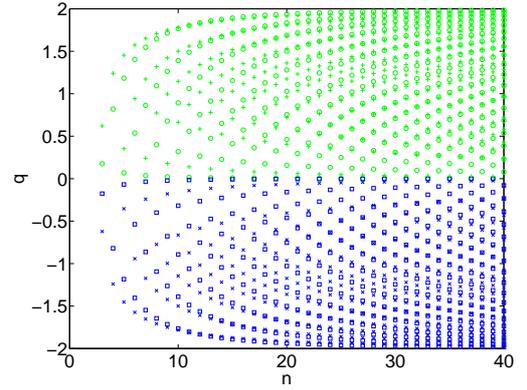}}
\caption{Frequencies of inertial modes in the co-rotating frame, $q$, 
for $\beta=0$ (circles and squares) and $\beta=18$ (x and plus) as a
function of the spatial degree $n$. Only the modes with $k=-1$ (green, respectively plus and circles) and $k=1$
(blue, respectively x and squares) are shown. }
\label{frequencies}
\end{figure}

We start by investigating the effect of compressibility on individual inertial
modes and the growth rate of triad resonances. We will first consider inviscid
flows and include viscous effects at the end of this section.

Fig. \ref{frequencies} shows the dispersion relation for $k=-1$ and $k=1$ for
incompressible fluids and for $\beta=18$. As expected, compressibility modifies
the inertial mode frequencies. There is a noteworthy distinction between the
cases $kq>0$ and $kq<0$ in that the modes with the frequency $q$ with the
smallest absolute value at any given $n$ always occurs for $kq<0$. These low
frequency modes will become relevant later on.

The mode frequencies change
continuously with $\beta$ so that individual modes can be tracked as a function
of $\beta$. Fig. \ref{sigbeta1} and \ref{sigbeta2} shows a few examples of triads with
$(k_i,k_j)=(-1,1)$. It can be seen that a chosen triad may lead to instability
or not depending on $\beta$. A variation of $\beta$ modifies both the structure
and frequency of the inertial modes and hence the integrals in $V_{i,j}$ and
$\delta \omega$ in Eq. (\ref{gledz1}). As expected from this equation, the
positive growth rates in figs. \ref{sigbeta1} a) and \ref{sigbeta2} a) are found around the
$\beta$ for which $\delta \omega = 0$. Figures \ref{sigbeta1} b) and \ref{sigbeta2} b) give
the variation of $\delta \omega$ with $\beta$ and demonstrate the effect of
compressibility on mode frequencies.

The case $|k|=1$, $n=2$ is exceptional because it corresponds to the purely toroidal
motion in the spin-over mode. All toroidal modes have no radial velocity
component, which implies for a radially dependent background density profile
$\rho_0$ that $\rho_0$ drops from the continuity equation
$\nabla\cdot(\rho_0\mathbf u)=0$. The density profile $\rho_0$ then disappears
completely from the eigenvalue and stability problems, so that $\beta$ does not
affect the growth rate or the $\delta \omega$ in the triad $(-1,1)$ with $n=2$.
In addition, these two modes are the only ones in exact resonance with $\delta
\omega = 0$ at $\beta = 0$.
\begin{figure}
\centering
\subfigure[$\epsilon=1\cdot10^{-2}$, $Ek=0$ \label{epsi5e-3ek0} \newline ]{\includegraphics[width=0.4\textwidth,]{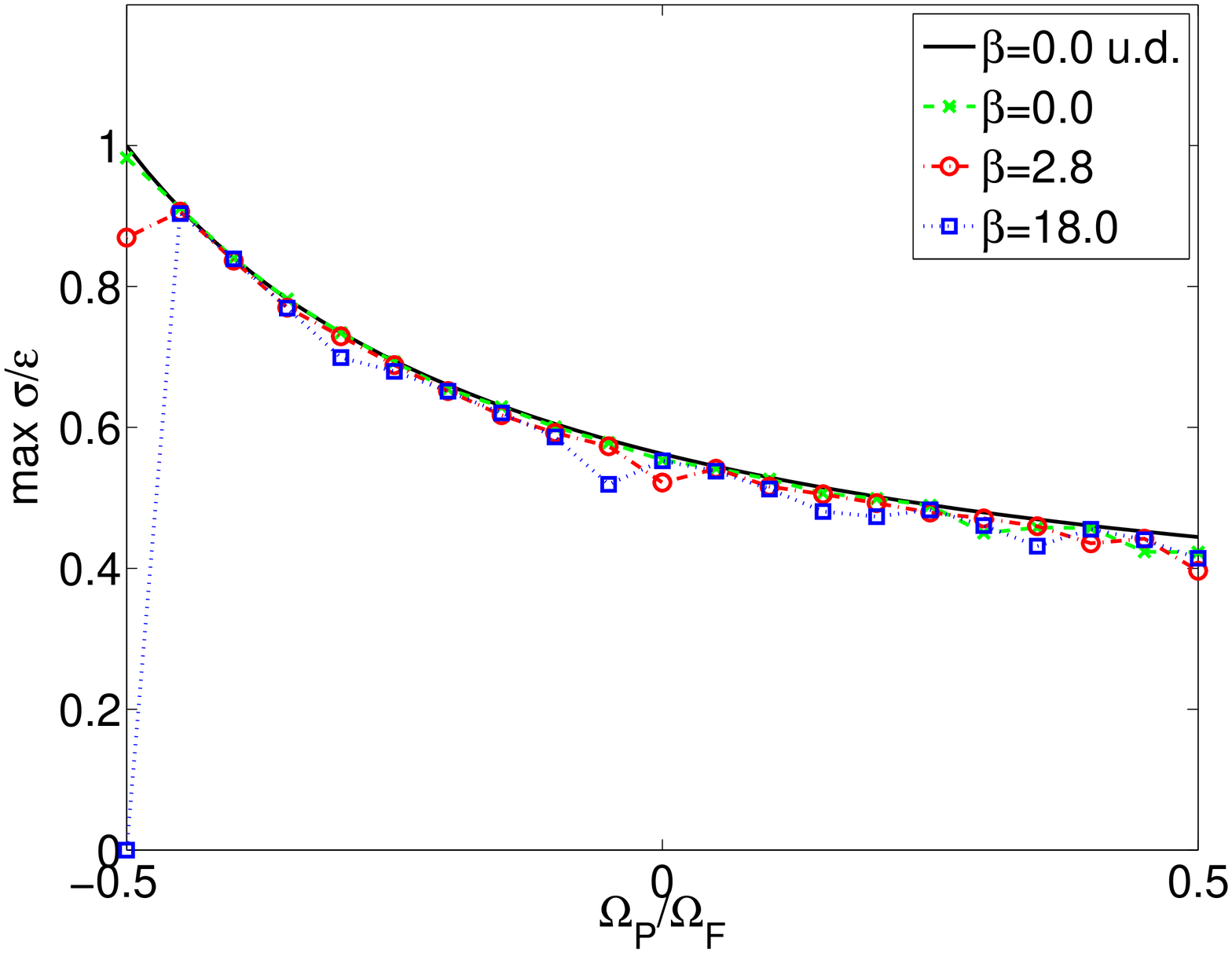}}
\subfigure[$\epsilon=1\cdot10^{-3}$, $Ek=0$ \label{epsi5e-4ek0}\newline ]{\includegraphics[width=0.4\textwidth]{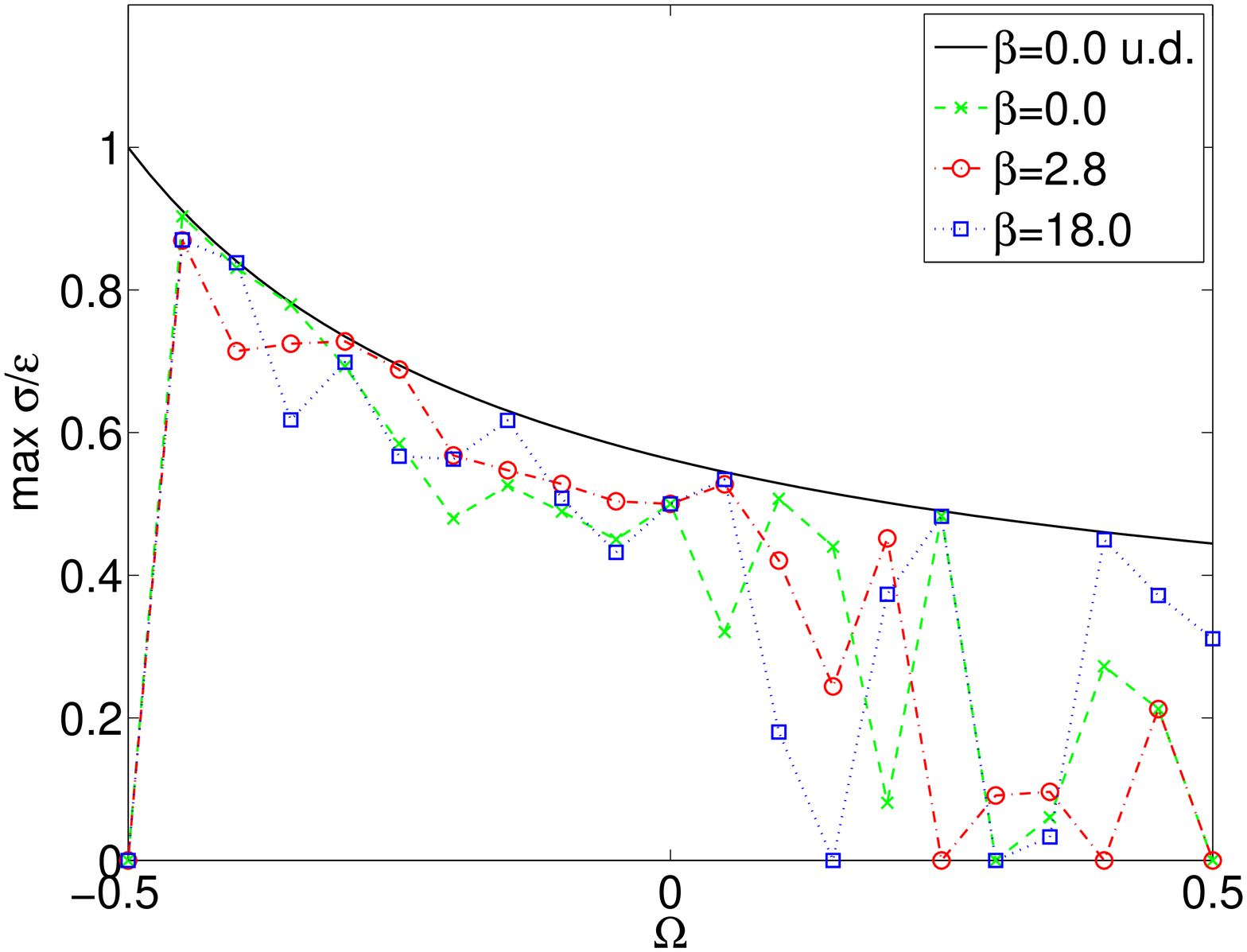}}
\caption{Maximum growth rates as a function of $\Omega$ for Ek=0. For this figure we take
into account all modes with $|k| \le 10$ and $n \le 40$. The black (solid) line in this figure is the growth rate given by Eq. (\ref{sigmaud})}
\label{sigmax}
\end{figure}
It will now be argued that  while the properties of inertial modes depend on
compressibility, the stability of the flow as a whole does not. For an inviscid
fluid, the two control parameters apart from $\beta$ determining stability are
the ellipticity of the deformation, $\epsilon$, and the ratio of the rotation
rates $\Omega$ defined below Eq. (\ref{gledz1}). The tidal flow is unstable if any two inertial
modes form a triad with positive growth rate. Since there is an infinity of
modes, a computational search for instability is necessarily restricted to a
finite subset. Figures \ref{sigmax} a) and b) are the result of a scan of all
modes with $|k| \leq 10$ and $n \leq 40$. These two panels display the maximum
growth rate found in this subset for the chosen $\epsilon$ as a function of
$\Omega$ for different $\beta$. Inertial modes appear only within a limited
band of frequencies, so that resonances are impossible in the interval $-3/2 <
\Omega < -1/2$ \citep{craik1989}. Fig. \ref{sigmax} shows a part of the
complementary interval by way of example.  The various curves appear to have a
common envelope $\sigma_{ud}$ given by
\begin{equation}
 \sigma_{ud}=\frac{(3+2\Omega)^2}{16(1+\Omega)^2} \epsilon.
\label{sigmaud}
\end{equation}
This expression stems from the calculation of the inviscid growth rate in an
incompressible, infinitely extended fluid \citep{miyazaki1995}. Some deviations
remain between Eq. (\ref{sigmaud}) and the computed maximum growth rates in
fig. \ref{sigmax}. However, these deviations become smaller and smaller if
larger numbers of eigenmodes are included in the search for unstable triads.
This can already be seen by comparing figures \ref{sigmax} a) and b): Figure
\ref{sigmax}b) is for the smaller $\epsilon$. For a given subset of allowed
eigenmodes, fewer pairs of modes will meet the criterion $|\delta\omega|\leq
O(\epsilon)$ required by Eq. (\ref{gledz1}) for the smaller $\epsilon$,
so that larger deviations remain between the envelope and the computed
maximum growth rates. It is plausible to assume that the deviations will entirely
disappear if all inertial modes are taken into consideration.

We thus arrive at the picture that the stability limit does not depend on
$\beta$, but $\beta$ determines which modes are unstable. We will now compute
the spatial degree $n$ of the most unstable modes. A resonance can only occur
if two modes $i$ and $j$ have frequencies such that $\omega_i-\omega_j <
O(\epsilon)$. 
For a smaller $\epsilon$, we expect for purely statistical reasons that
a larger pool of eigenmodes is
necessary to find any resonance and that the unstable modes have a
higher $n$ for smaller $\epsilon$. Because $\omega_i=(1+\Omega)q_i-k_i$ for all
$i$, the frequencies $q_i$ and $q_j$ have to obey
\begin{equation}
 q_i-q_j=\frac{k_i-k_j}{1+\Omega} + \frac{1}{1+\Omega} O(\epsilon).
\label{q_resonance}
\end{equation}
In other words, if there is a mode at frequency $q_j$, it can only form an
unstable triad if there is another mode with a frequency in an interval of size
$O(\epsilon)/(1+\Omega)$ centered around $q_j+(k_i-k_j)/(1+\Omega)$. Let us
first consider the case of small $|\Omega|$, so that $q_i-q_j \approx k_i-k_j +
O(\epsilon)$. All eigenmode frequencies obey $|q_i| \le 2$. 
Assuming that these frequencies 
are statistically uniformly distributed over the interval available for inertial
modes, and that the matrix elements appearing in Eq. (\ref{gledz1}) do not
systematically vary with frequency, one will find on average one resonance with
positive growth rate among a number $N$ of modes proportional to $1/\epsilon$.
Only a finite number of modes exist with an $n$ smaller than some limit $L$: For
every $n$ and $k$, there are $n-|k|$ modes if $k \neq 0$ and $n-1$ modes if
$k=0$. For every $n$, the index $k$ has to obey $|k| \leq n$ so that there are
$n^2-1$ modes of spatial degree $n$ and $N(L)$ modes exist with $n \leq L$,
with $N(L)$ given by:
\begin{equation}
 N(L)=\sum_{n=1}^L (n^2-1)=\frac{1}{6}L(L+1)(2L+1)-L.
\label{totalnumber}
\end{equation}
For large $L$, one has $N \propto L^3$, and in order for
these modes to contain a resonance, one needs $L \propto \epsilon^{-1/3}$,
independently of $\Omega$ or $\beta$. The
smaller the deformation $\epsilon$, the higher the typical spatial complexity
of the modes involved in resonances.

\begin{figure}[h!tbp]
\begin{center}
\includegraphics [width=0.4\textwidth]{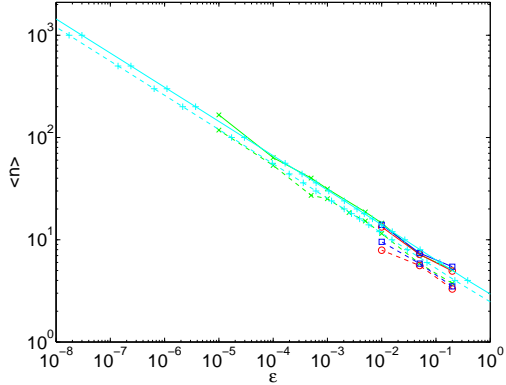}
\caption{Mean value of the spatial degree $<n>$, for the modes with a growth
rate of at least $0.1\sigma_{ud}$ (dashed lines) or $0.8\sigma_{ud}$
(solid lines) as a function of $\epsilon$ for $\beta=0$ (green x),
$\beta=2.8$ (red circles), and $\beta=18$ (blue squares), together with fits according to
Eq.\ref{totalnumber}.}
\label{eklepsi}
\end{center}
\end{figure}

\begin{figure}
\centering
\subfigure[\label{orbene0} \newline ]{\includegraphics[width=0.4\textwidth,]{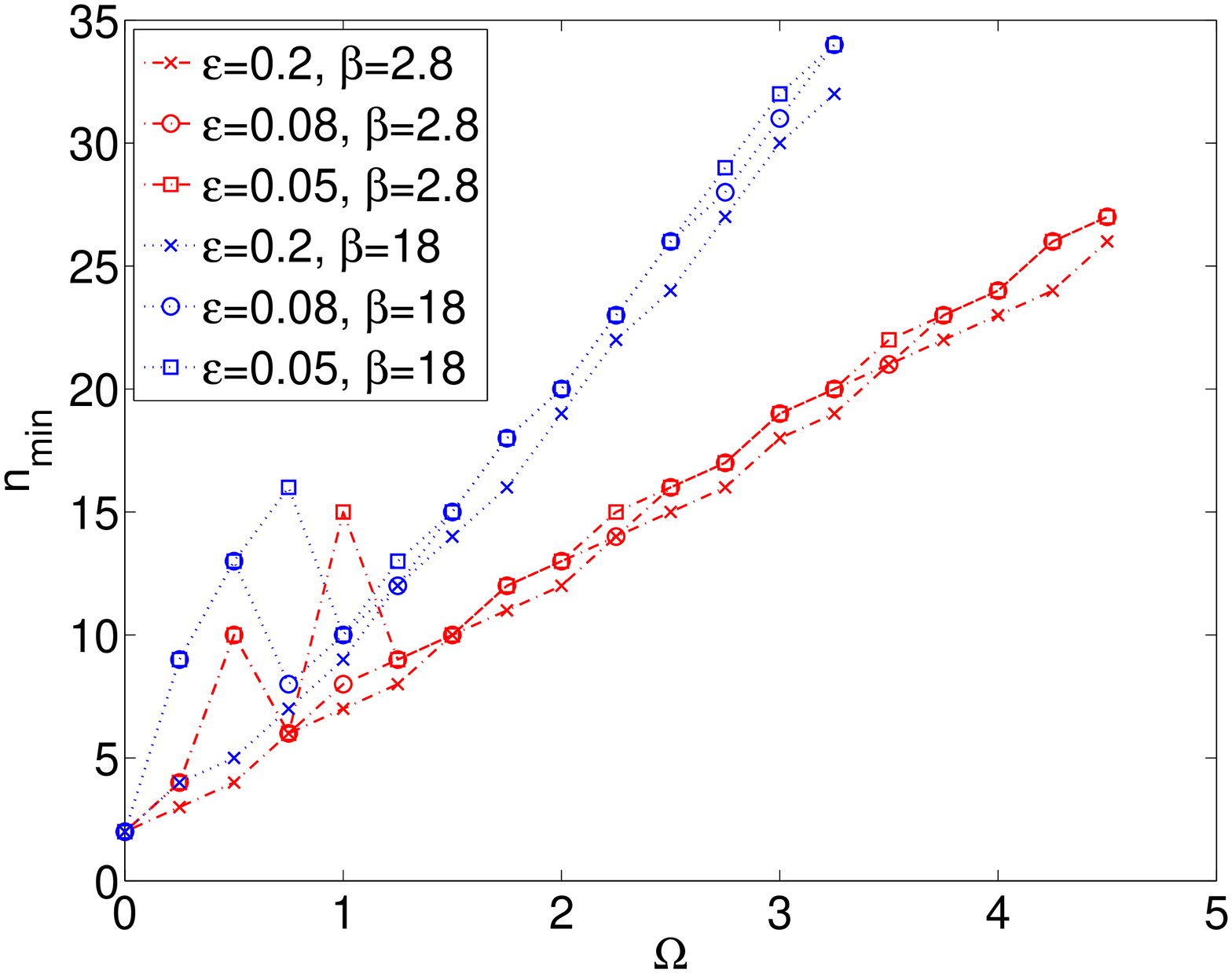}}
\subfigure[\label{orbeeq0}\newline ]{\includegraphics[width=0.4\textwidth]{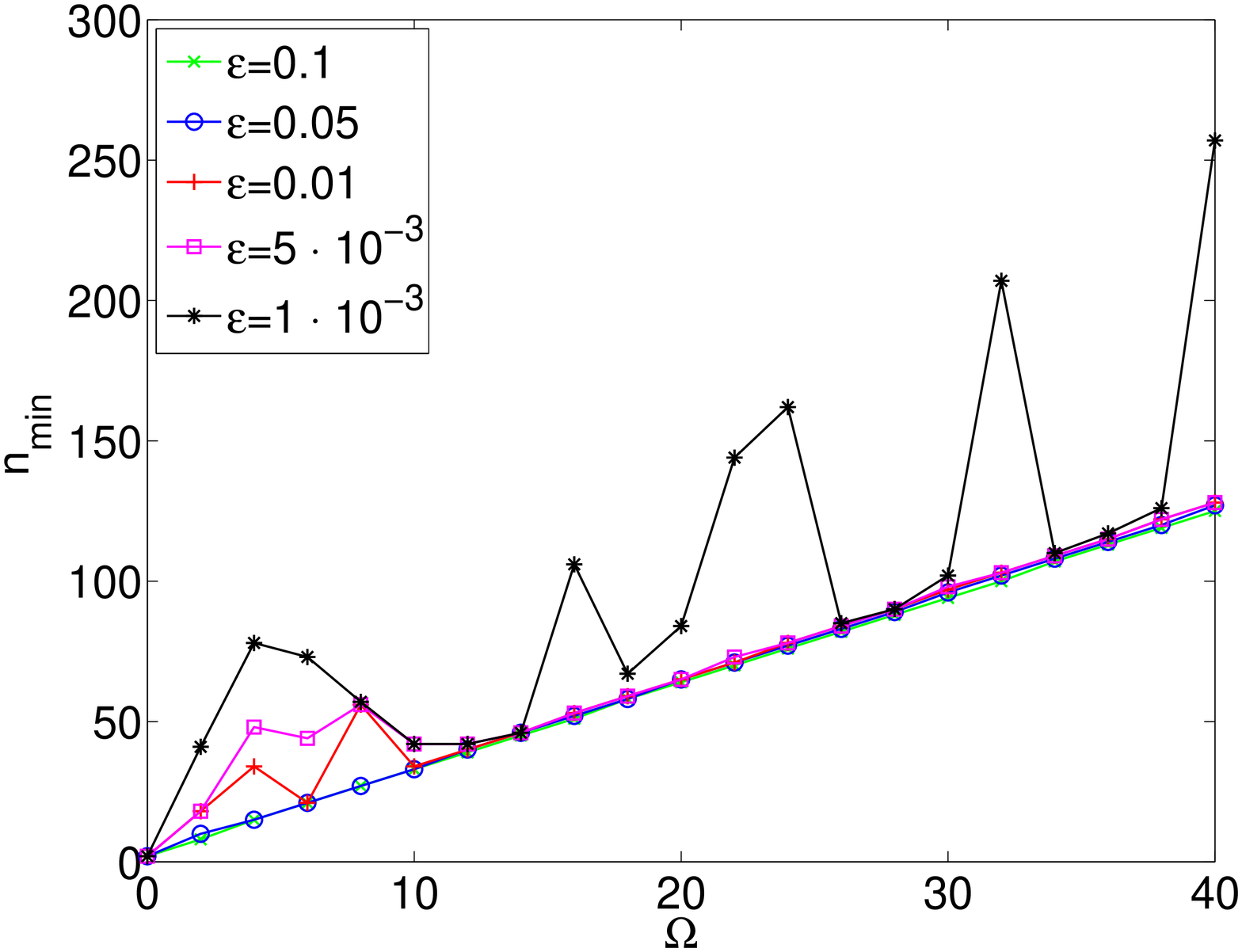}}
\caption{Minimum value of $n$, $n_{min}$ as a function of $\Omega$ for various
$\epsilon$ and positive $\Omega$. (a) $\beta=2.8$ and $\beta=18$ and (b)  $\beta=0$.
}
\label{lfepsi}
\end{figure}

\begin{figure}
\centering
\subfigure[\label{orbene0neg} \newline ]{\includegraphics[width=0.4\textwidth,]{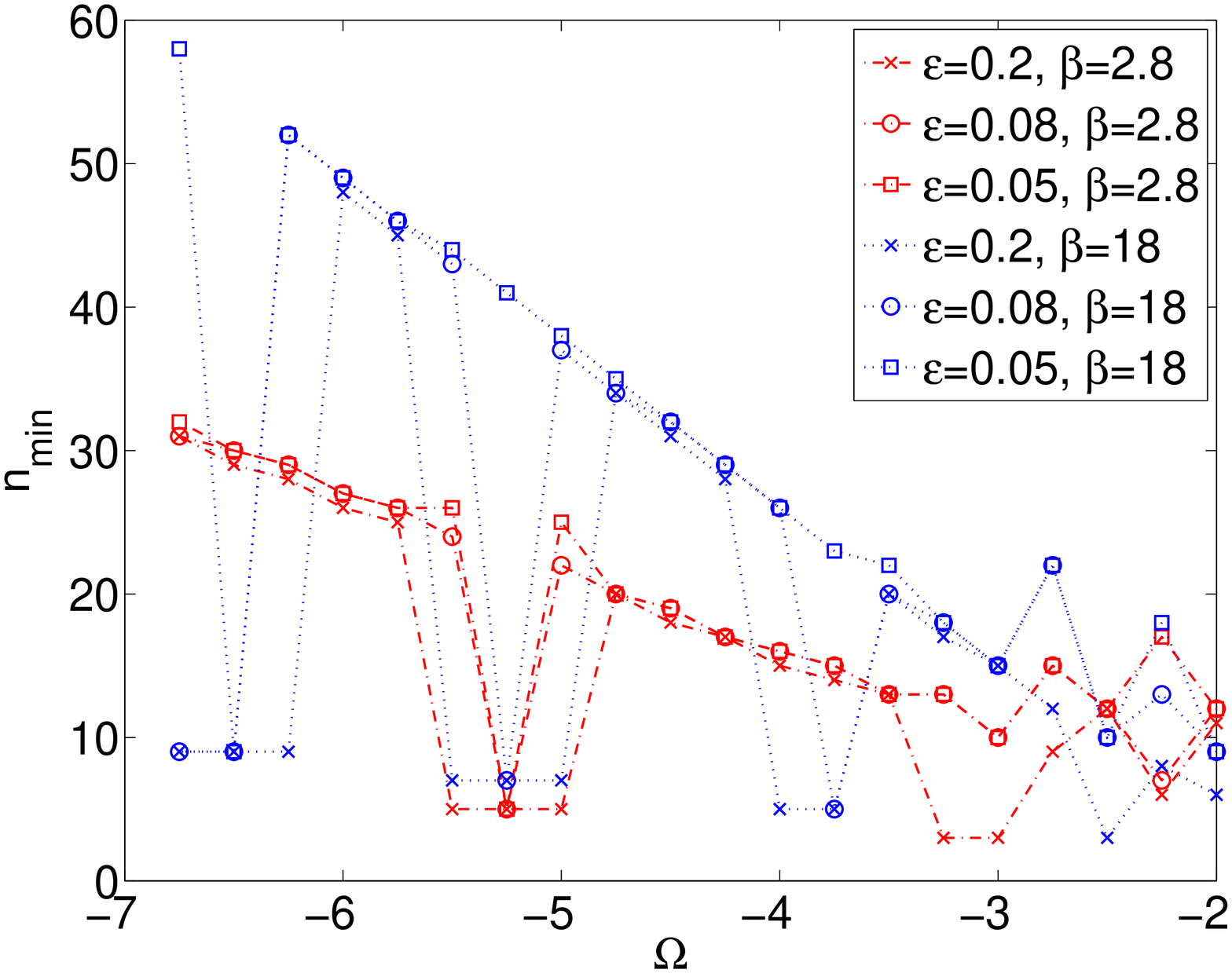}}
\subfigure[\label{orbeeq0neg}\newline ]{\includegraphics[width=0.4\textwidth]{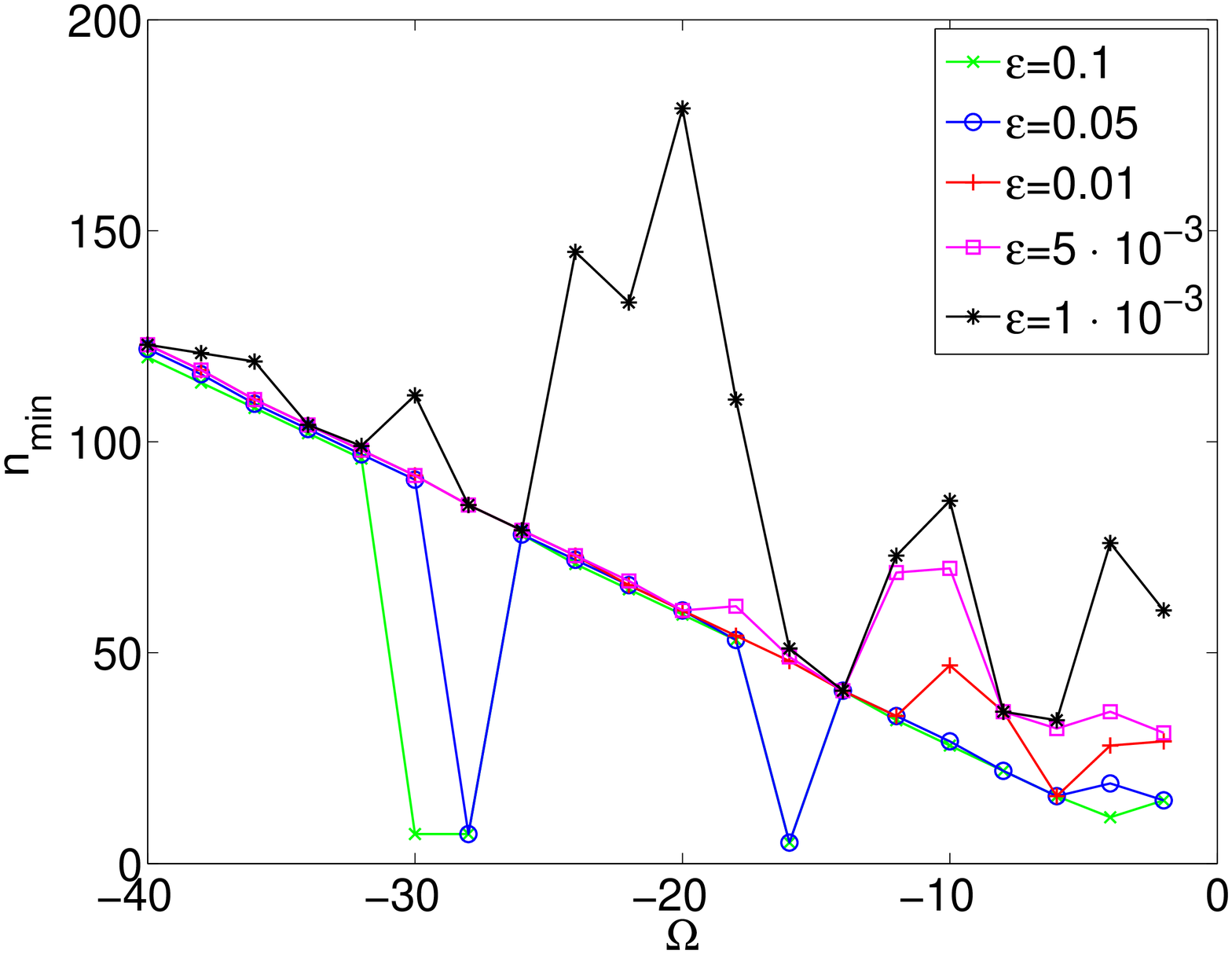}}
\caption{
Same as in fig. \ref{lfepsi} but for negative $\Omega$.
}
\label{nmin-neg-om}
\end{figure}

Fig. \ref{eklepsi} verifies this scaling. This figure has been obtained as
follows: At fixed $\epsilon$, 20 equidistant points have been chosen in
the interval of $\Omega$ ranging from $-0.45$ to $0.5$. For every point, we 
determined the triad with the mode with the smallest $n$ for which the growth
rate is at least $0.1 \sigma_{ud}$ and  $0.8 \sigma_{ud}$. This minimal $n$,
averaged over all points,
is $<n>$ and is shown in figure \ref{eklepsi} as a function of $\epsilon$.
The averaging procedure is intended to remove the occasional outlier which occurs
if a mode with a very low $n$ happens to be part of a growing triad. The average
$<n>$ is rather a measure of the minimum $n$ that is typically necessary to
obtain a resonance with a given $\epsilon$. As expected from the argument above,
$<n> \propto \epsilon^{-1/3}$ for small $\epsilon$.

We now turn to the case of large $|\Omega|$ in Eq. (\ref{q_resonance}). For
$|\Omega|$ tending to infinity, $q_i-q_j$ tends to zero. At the same time, we see
from Eq. (\ref{gledz1}) that positive growth rates are only possible for $q_iq_j
< 0$, so that $q_i$ and $q_j$ both have to go to zero as $1/|\Omega|$ if
$|\Omega|$
goes to infinity. Fig. \ref{frequencies} shows that $n$ must be large enough to
find any eigenvalue with an absolute value below some prescribed bound. This yields a
minimal spatial degree $n_{min}$ necessary to find a resonance at a given
$\Omega$. The corresponding mode frequency needs to lie in the appropriate
interval of size $O(\epsilon)/(1+\Omega)$ for a positive growth rate. This
additional condition is possibly met only for $n$ larger than $n_{min}$.

We therefore need a relationship between the smallest frequencies and the spatial degrees of the modes. 
As already stated the frequencies of inertial modes in a uniform density sphere
can be calculated according to Eq. (29) from \citet{wu2004}
, see Eq.(\ref{freq}).

For large $n$ the asymptotic expansion  \citep{abramowitz1972}
\begin{equation}
\begin{split}
P_n^k(\cos\theta)=\frac{\Gamma(n+k+1)}{\Gamma(n+3/2)}\left(\frac 1 2 \pi \sin(\theta) \right)^{-1/2} \\ \cdot \cos\left[\left(n+\frac 1 2\right)\theta-\frac \pi 4 +\frac{k\pi}{2}\right]+O(n^{-1}) 
\end{split}
\end{equation}
holds. We set $\mu=\cos(\theta)$ and $\theta=\pi/2+\epsilon_w$ with $\epsilon_w$
small for $\mu$ small. A short calculation shows that for small $\mu$ and large n, (\ref{freq}) is equivalent to 
\begin{equation}\label{frerel}
\begin{split}
-k\cos\left[\left(n+\frac 1
2\right)\epsilon_w+\frac{(n+k)\pi}{2}\right]\\=\left(n+\frac 1
2\right)\sin\left[\left(n+\frac 1 2\right)\epsilon_w+\frac{\left(n+k\right)\pi}{2}\right].
\end{split}
\end{equation}
For $n+k$ even this gives 
\begin{subequations}
 \begin{align}
  \epsilon_w&\approx \frac{-k}{\left(n+\frac 1 2\right)^2} \quad \mbox{for} \quad \left(n+\frac 1 2\right)\epsilon_w\rightarrow 0 \\
  \epsilon_w&\approx \pm\frac{ h\pi}{n+\frac 1 2} \quad \mbox{for} \quad \left(n+\frac 1 2\right)\epsilon_w\rightarrow \pm h\pi
 \end{align}
\end{subequations}


and for $n+k$ odd 
\begin{equation}
 \epsilon_w\approx  \pm\frac{ h\pi}{2\left(n+\frac 1 2\right)} \quad \mbox{for} \quad  \left(n+\frac 1 2\right)\epsilon_w\rightarrow \pm \frac{h\pi}{2},
\end{equation}
where $h$ is an integer. For large $n$ and small
k the $\mu$ with the smallest absolute value tends towards zero as
$k/(n+1/2)^2$. We can classify these modes as ``slow modes'', because if the
frequency of modes on any other branch of the dispersion relation in fig.
\ref{frequencies} tends to zero, it does so only in $1/(n+1/2)$. We
thus have to distinguish two cases when computing $n_{min}$: the growing
resonance is either between two slow modes, or at least one of the two inertial
modes involved is not a slow mode. In the latter case, $n_{min}\propto
|\Omega|$, whereas in the former case, $n_{min}\propto \sqrt{|\Omega|}$.
However, since $|k_1-k_2|=2$, and choosing the indices such that $k_2>k_1$, a
resonance between slow modes can only occur if $k_1=-1$ and $k_2=1$, and from
(\ref{q_resonance}) we can deduce that $\Omega$ must be negative.
In summary, we expect $n_{min}\propto\Omega$ for positive $\Omega$, and for
negative $\Omega$, too, except when modes with frequencies of approximately $k/(n+1/2)^2$ can resonate with each other.

\begin{figure}
\centering
\subfigure[$\epsilon=1\cdot10^{-2}$, $Ek=1\cdot10^{-5}$
\label{epsi0,01dsigtek1e-5}]{\includegraphics[width=0.4\textwidth,]{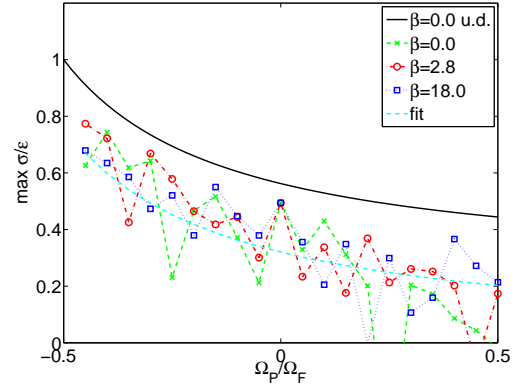}}\\ 
\subfigure[$\epsilon=1\cdot10^{-2}$,
$Ek=1\cdot10^{-6}$\label{epsi5e-3ek4e-8tasso}]{\includegraphics[width=0.4\textwidth,]{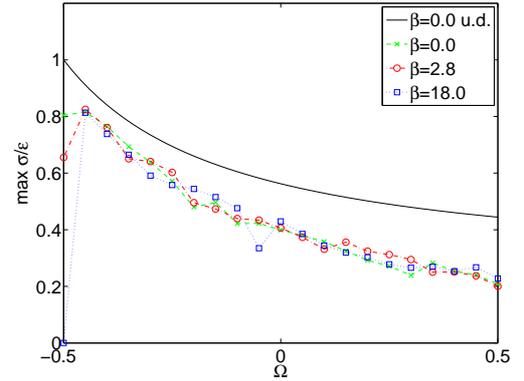}}
\caption{Maximum growth rates as a function of $\Omega$ for $Ek\ne 0$. For this figure we take
into account all modes with a maximum $|k| \le 10$ and $n \le 40$. (a) Calculated
for free slip boundary conditions according to Eq. (\ref{fsbc}). (b) Calculated
for no slip boundary conditions according to Eq. (\ref{nsbc}). The black (solid) line  in this figure is the growth rate given by Eq. (\ref{sigmaud}). The cyan (dashed without marker) line in (a) is the growth rate given by Eq. (\ref{fsbcb}) with $d_2=3.35$}
\label{sigmaxek}
\end{figure}

Fig. \ref{lfepsi} proves this scaling to be correct for positive $\Omega$. We
find $n_{min}=d_1 \Omega$, with a prefactor $d_1$ which depends on $\beta$. We
can deduce from that figure that $d_1=3$, 5, and 10 for $\beta=0$, $2.8$, and 18,
respectively. Some points lie above the line $n_{min}=d_1 \Omega$ because some
modes with spatial degree $n_{min}$, even though their frequency is small
enough, do not find another mode to resonate with. For negative $\Omega$ (fig.
\ref{nmin-neg-om}), one also finds points below $n_{min}=d_1 |\Omega|$. These
correspond to resonances between two slow modes and always involve the azimuthal
wavenumbers $k_1=-1$ and $k_2=1$. Since $\Omega >0$ in most applications, we
will use $n_{min}=d_1 \Omega$ in the estimation of dissipation effects below.


We have up to here considered ideal fluids only. The remainder of this section 
will deal with viscous effects.
Fig. \ref{sigmaxek} shows the maximum growth rate if dissipation is
heuristically taken into account according to equations (\ref{fsbc}) and
(\ref{nsbc}). The dissipation for no slip boundaries formally depends on $n$.
However, at the $Ek$ and $n$ under consideration, the first term corresponding
to friction at boundaries dominates and it is independent of mode structure.
Therefore, the maximum growth rates
in figure \ref{sigmax} are the same as before except for a
downward shift, and are given by $\sigma_{ud}-g_2 |1+\Omega| \sqrt{Ek}$.
Instability is expected if 
$\sigma_{ud}-g_2 |1+\Omega| \sqrt{Ek} > 0$, independently of $\beta$. Values of
$c$ depend on the inertial modes and may vary by an order of magnitude, but
since the uncertainties on $Ek$ are in practice much larger, there is little
incentive to determine the prefactor $c$ more accurately.

Free slip boundary conditions allow only bulk dissipation and are more
complicated because dissipation now increases with $n$. This puts a natural limit to
the spatial degrees $n$ it is useful to consider, which restricts altogether the
number of modes suitable for instabilities because $|k| \leq n$ and there is
only a finite number of modes for a given $k$ and $n$. If one includes more and
more eigenvalues in the stability analysis, the curve of the maximum growth rate
approaches a smooth curve of the same type as $\sigma_{ud}$ up to some
approximation, until additional modes suffer from such a high damping rate that
they can not improve the maximum growth rate any more.

The precise stability limit now depends on the structure of the unstable modes
and hence on $\beta$ and $\Omega$, but as an estimate, we can take the growth rate
for free slip boundaries to be the inviscid growth rate reduced by the decay
rate of modes with the expected spatial degree, for which we may reasonably take $<n>$
for small $|\Omega|$ and $n_{min}$ for large $|\Omega|$, or as a formula
interpolating between these two limits, $n=d_1 |\Omega|+ d_2 \epsilon^{-1/3}$. For
the parameters of figure \ref{sigmaxek}, only the term in $d_2 \epsilon^{-1/3}$ is
important and one arrives at 
\begin{equation}\label{fsbcb}
\sigma_{fs}=\sigma_{ud} -  d_2^2 \epsilon^{-2/3} Ek,
\end{equation}
which is again independent of
$\beta$. From figure \ref{epsi0,01dsigtek1e-5}, we
deduce that $d_2$ is roughly $3.35$.


\section{Examples:  Io's tides on Jupiter, the binary system V636 Centauri and the Earth}

The results from  the previous section are used  for three examples.  We
consider the $0.85 M_\odot$ secondary as the perturbing object within the binary
system V636 Centauri, the terrestrial core in the Earth Moon system, and the tides
on Jupiter raised by Io.

The choice of the correct boundary conditions is generally ambiguous.
For the Earth's core it is clear that no-slip boundary conditions apply. 
The choice for the two remaining examples is not so clear. One finds in the
literature arguments both in favor of no slip (\citet{tassoul1987,tassoul1995},
see also \citet{tassoul1997}) as well as free slip boundary conditions
(\citet{rieutord1992,rieutord2008}, see also \citet{rieutord1997}).

An even bigger problem is how to deal with additional fluid motion, such as
convection, on which the tides are superimposed. The simplest approach is to
ignore such motions, which is certainly justified if their amplitudes is much
smaller than the amplitude of the tidal flow. Another approach consists in
introducing a turbulent viscosity. However, turbulence modeling is always
uncertain, and according to at least one well established turbulence model,
elliptical instability is enhanced and not suppressed under some circumstances
\citep{fabijonas2003}. When using a turbulent viscosity for the present problem, one has to deal with
the unusual situation that one needs to compute the damping of a motion with a
period which is typically shorter than the turn over time of the turbulent
eddies. Several authors (\citet{wu2004b} and in a similar form \citet{ogilvie2007}) have used the following expression for the
turbulent viscosity $\nu_t$
\begin{equation}
 \nu_t\sim v_{cv}l_{cv}\frac{1}{1+(\omega_{tide}\tau_{cv}/(2\pi))^{s_e}},
\label{nu_t}
\end{equation}
where $v_{cv}$, $l_{cv}$, and $\tau_{cv}$ are the characteristic convection
velocity, mixing length, and turnover time and $\omega_{tide}=2|\Omega_F|$ is the frequency of the tidal forcing. $s_e$ is a constant, generally $s_e=1$ or
$s_e=2$ is used. \citet{goldreich1977} and \citet{goodman1997} suggested that
$s_e=2$, while \citet{zahn1977} argues for $s_e=1$. For tides faster than
$\tau_{cv}$, one expects a turbulent
viscosity proportional to $v_{cv}$ and a mixing length given by the distance
traveled by particles during one tidal period, $l_{cv} T_f/\tau_{cv}$, with $T_f=2\pi/\omega_{tide}$, which corresponds to $s_e=1$ in Eq. (\ref{nu_t}). Numerical simulations
by \citet{penev2009} confirm this expectation, but because of limited spatial
resolution they were not able to simulate reliably the case of $T_f \ll
\tau_{cv}$, the regime for which $s_e=2$ has been proposed. 

That $s_e=2$ should be used at small tidal periods is in accordance with results from \citet{ogilvie2012}.
Therfore we use $s_e=2$ for the case $T_f\ll \tau_{cv}$ and $s_e=1$ for $T_f\gtrsim \tau_{cv}$. In order to apply Eq. (\ref{nu_t}), one needs
numbers for $v_{cv}$, $l_{cv}$, and $\tau_{cv}$. The mixing length is estimated
by $l_{cv}\approx H$, with $H=-\frac{d r}{d \ln \rho}=\frac{r^{-1}-r}{2 \beta}$
the density scale height. Therefore an order of magnitude estimation gives
$l_{cv}\sim\frac {R_c}{2\beta}$. 
The convective velocity is approximated by 
$v_{cv}\approx(\textrm{F}/\rho)^{1/3}$, with the energy flux $\textrm{F}=L/(4\pi R_c^2)$ and $R_c$ the radius of the central body. 
An order of magnitude for $v_{cv}$ is therefore obtained by using
$v_{cv}\approx(3 R_c^3 \ \textrm{F}/(4m_c))^{1/3}$, with $m_c$ the mass of the central body. Finally,
$\tau_{cv}=l_{cv}/v_{cv}$.

Table \ref{allv} lists for the three examples the growth rate for an inviscid
fluid, and the damping rates computed for molecular and turbulent viscosities.
In the no slip case the damping rate $D$ is given by $D=(1+\Omega)\sqrt{Ek}$
according to Eq. \ref{nsbc}. We neglect the volume damping term because
$\Omega$ and $Ek$ are small enough in both cases. In the free slip case,
$D=d_2^2 \epsilon^{-2/3}Ek$ according to Eq. \ref{fsbcb}. For $d_2$ we choose $3.35$
, the value extracted from figure \ref{epsi0,01dsigtek1e-5}. We neglect the terms which
are relevant if $\Omega$ is large (see the last paragraph in section 4) because $\Omega$ is small enough in both cases.
The ellipticity is calculated according to
\begin{equation}
 \epsilon=\frac{1}{2}\frac{m_p}{m_c}\left(\frac{R_c}{a_{cp}}\right)^3 
\end{equation}
with $m_p$ the mass of the perturbing body and $a_{cp}$ the distance between the central and the perturbing body. The Ekman number is calculated according to
\begin{equation}
 Ek=\frac{\nu}{\Omega_{spin}R_E^2}
\end{equation}
with $R_E=R_{co}$ the radius at the edge of the outer conducting region in the case of the Earth and the Jupiter-Io system and $R_E=R_c$ in the case V636 Centauri and $\Omega_{spin}=\Omega_F+\Omega_P$.

Instability is expected if the inviscid growth rate exceeds the viscous damping
in table \ref{allv}. Table \ref{twocase} summarizes the result of different
modeling assumptions. In practice, the turbulence model decides on whether
instability is predicted or not. Unfortunately, we have no turbulence model we
can safely rely upon. 

\begin{table}
\caption{Parameters for the example objects.}
\label{allv}
\centering
\begin{tabular}{c|ccc}
 Constant  & Earth-Moon & Jupiter-Io & V636 Cen. \\
\hline
$R_{c} (m)$&$6.37\cdot 10^6$ \tablefootmark{a}&$6.99\cdot10^7$ \tablefootmark{a}&$7.08\cdot10^8$ \tablefootmark{b}\\
$R_{co} (m)$&$3.47\cdot10^6$ \tablefootmark{c}&$5.71\cdot10^7$ \tablefootmark{c}&...\\
$m_{c} (kg)$&$5.97\cdot 10^{24}$ \tablefootmark{c}&$1.90\cdot10^{27}$ \tablefootmark{c}&$2.09\cdot10^{30}$ \tablefootmark{b}\\
$a_{cp} (m)$&$3.84\cdot 10^8$ \tablefootmark{c}&$4.22\cdot10^8$ \tablefootmark{c}&$9.57\cdot 10^9$ 
\tablefootmark{d}\\
$m_{p}(kg)$&$7.35\cdot10^{22}$\tablefootmark{c}&$8.91\cdot10^{22}$ \tablefootmark{c}&$1.70\cdot10^{30}$ \tablefootmark{b}\\
$L (W)$&...&$3.34\cdot10^{17}$ \tablefootmark{e}& $4.31\cdot10^{26}$ \tablefootmark{b}\\
$\nu_m (m^2s^{-1})$ & $1.4\cdot10^{-6}$ \tablefootmark{f}&$3\cdot10^{-7}$ \tablefootmark{e}&$10^{-4}$ \tablefootmark{g}\\
$\Omega_{spin}(\textrm{day}^{-1})$&$2\pi $&$2\pi/0.41 $ \tablefootmark{h}&$2\pi/3.96$ 
 \tablefootmark{i} \\
$\Omega_{P}(\textrm{day}^{-1})$&$2\pi/27.32 $&$2\pi/1.77$ \tablefootmark{h}&$2\pi/4.28$ \tablefootmark{b}\\
$\Omega$&0.0380&0.305&12.40\\
$\epsilon$&$2.8\cdot10^{-8}$&$1.1\cdot10^{-7}$&$1.6\cdot10^{-4}$\\
$E_m$&$1.6\cdot10^{-15}$&$5.2\cdot 10^{-19}$&$1.1\cdot10^{-17}$\\
$\beta$&...&1.5&18\\
$s_e$&...&2&1\\ 
$E_{turb}$&...&$1.6\cdot10^{-13}$&$6.3\cdot10^{-5}$\\
$\sigma_{ud}$&$ 1.6\cdot10^{-8}$&$5.1\cdot10^{-8}$&$4.4\cdot10^{-5}$\\
$D_{m}$ n.s.&$ 4.2 \cdot 10^{-8} $&$9.4\cdot10^{-10}$&$4.4 \cdot 10^{-8}$\\
$D_{m}$ f.s.&...&$2.6\cdot 10^{-13} $& $4.1\cdot10^{-14}$\\
$D_{turb}$ n.s.&...&$5.3\cdot10^{-7}$&$0.11$\\
$D_{turb}$ f.s.&...&$8.3\cdot10^{-8}$&$0.24$
 \end{tabular}
 \tablefoot{$\nu_m$ is the molecular
viscosity, $E_m$ the Ekman number based only on the molecular viscosity, and
$E_{turb}$ the Ekman number based on the turbulent viscosity (the molecular is
negligible in comparison to the turbulent in these examples). $D_m$ and $D_{turb}$ are damping constants based on the molecular viscosity and the turbulent viscosity, n.s. and f.s stands for no slip and free slip boundary conditions.\\
   \tablefoottext{a} {\citet{archinal2010}}
   \tablefoottext{b} {\citet{clausen2009}}
    \tablefoottext{c} {\citet{wicht2010}}
    \tablefoottext{d} {Calculated according to Kepler's third law, with values given in \citet{clausen2009}}
    \tablefoottext{e} {\citet{guillot2004}}
    \tablefoottext{f} {\citet{dewijs1998}}
    \tablefoottext{g} {\citet{miesch2005}}
    \tablefoottext{h} {\citet{wu2004b}}
    \tablefoottext{i} {Calculated  according to $v=\Omega_{spin}\cdot R_c$ with $v$ the equatorial velocity  given in \citet{clausen2009}}
}
\end{table}

\begin{table}
\caption{Summary of the stability characteristics of the flow for the
examples Jupiter-Io (J-I) and V636 Centauri (VC).}
\label{twocase}
\centering
\begin{tabular}{c|cccc}
b.c. & J-I: $\nu_m$ & J-I: $\nu_t$ & VC: $\nu_m$ & VC: $\nu_t$   \\
\hline
n.s. & unstable & stable  & unstable & stable \\
f.s. & unstable &  uncertain & unstable & stable
 \end{tabular}
\tablefoot{The boundary conditions (b.c.) are given in the first column (n.s. stands for
no slip and f.s. for free slip) and the top row indicates if only molecular
($\nu_m$)    turbulent viscosity ($\nu_t$) has
been used in the computation of the viscous damping.}
\end{table}


\section{Conclusion and discussion}

We computed the linear stability limit of tidal flow within the anelastic
approximation through a perturbation calculation in which the small parameter is
the deformation of the central body from spherical shape. The instabilities are
described as superpositions of two inertial modes of the rotating sphere. The
perturbation calculation is tractable if the initially three dimensional problem
of the computation of the inertial modes is separable and reduces to the
solution of an ordinary differential equation. This is the case if the density
profile obeys a power law. For an inviscid fluid, the growth rate of the
combination of two chosen modes depends on the density profile, but the growth
rate maximized over all possible combinations does not. Furthermore, the maximum
growth rate is the same as the one known for elliptical instability in
infinitely extended, incompressible fluids. The fact that one finds the same
maximum growth rates in so widely different situations justifies some tolerance
towards modeling assumptions. For example, we restricted our analysis to density
profiles obeying power laws and to the boundary condition that the velocity
normal to the boundary be zero at the surface. After the calculations presented
here, we expect that different modes will be found to be the most unstable for
more realistic assumptions, but that the stability limit of the flow will stay the
same. In the same fashion, we do not expect the particular choice of the major axis
parallel to the rotation axis in Eq. (\ref{eq13}) to affect the stability limit.

We added viscous effects empirically. If friction at a solid boundary dominates
dissipation, the stability is easily determined as a function of the rotation
rates of the central body and the tidal companion, the tidal deformation, and
the Ekman number. If on the other hand bulk dissipation dominates, the
dissipation depends on the flow structure of the unstable modes and therefore on
all details of the model, in particular the density profile. However,
dissipation depends mostly on the spatial degree which on average obeys a simple scaling
law as a function of tidal deformation, so that an estimate of the stability
limit is still possible.

Viscous damping also depends on boundary conditions, and models of an object's
interior help to decide on which approximation to the viscous damping is most
accurate. For instance, the dissipation in Ekman layers usually dominates if a
solid core is present. However, if one wants to predict the stability limit of
a particular astrophysical object, the biggest uncertainty comes from a possible
turbulent viscosity. Progress in the calculation of tidal dissipation thus
mostly hinges on advances in the treatment of turbulence.

\begin{acknowledgements}
The authors acknowledges research funding by Deutsche Forschungsgemeinschaft (DFG) under grant SFB 963/1, project A5.
\end{acknowledgements}

\bibliographystyle{bib/aa}
\bibliography{bib/veroef_bib}

\end{document}